\documentclass[twocolumn,twoside,superscriptaddress,pra,aps]{revtex4}
\usepackage{amsmath,mathrsfs,amsbsy,color,graphicx,bm,amsthm,amsfonts}
\usepackage{units}
\usepackage{bbm}
\usepackage{color}

\newcommand{\idol}{\ensuremath{\mathbbm 1}}

\begin{document}

\title{Quantum deleting and cloning in a pseudo-unitary system}
\author{Yucheng Chen}
\affiliation{School of Physical Science and Technology, Ningbo University, Ningbo, 315211, China}
\author{Ming Gong}
\affiliation{Key Laboratory of Quantum Information, University of Science and Technology of China, CAS, Hefei, 230026, China}
\author{Peng Xue}
\affiliation{Beijing Computational Science Research Center, Beijing 100084, China}
\author{Haidong Yuan}
\affiliation{Department of Mechanical and Automation Engineering, The Chinese University of Hong Kong, Hong Kong}
\author{Chengjie Zhang}
\email{chengjie.zhang@gmail.com}
\affiliation{School of Physical Science and Technology, Ningbo University, Ningbo, 315211, China}
\affiliation{School of Physical Science and Technology, Soochow University, Suzhou, 215006, China}

\begin{abstract}
In conventional quantum  mechanics, quantum no-deleting and no-cloning theorems indicate that two different and nonorthogonal states cannot be perfectly and deterministically deleted and cloned, respectively. Here, we investigate the quantum deleting and cloning in a pseudo-unitary system. We first present a pseudo-Hermitian Hamiltonian with real eigenvalues in a two-qubit system. By using the pseudo-unitary operators generated from this pseudo-Hermitian Hamiltonian, we show that it is possible to delete and clone a class of two different and nonorthogonal states, and it can be generalized to arbitrary two different and nonorthogonal pure qubit states. Furthermore, state discrimination, which is strongly related to quantum no-cloning theorem, is also discussed. Last but not least, we simulate the pseudo-unitary operators in conventional quantum mechanics with post-selection, and obtain the success probability of simulations. Pseudo-unitary operators are implemented with a limited efficiency due to the post-selections. Thus, the success probabilities of deleting and cloning in the simulation by conventional quantum  mechanics are less than unity,  which maintain the quantum no-deleting and no-cloning theorems.
\end{abstract}
\date{\today}


\maketitle

\section{Introduction}
Quantum no-deleting and no-cloning theorems are fundamental for quantum information science, which states that deleting and  creating identical copies of an arbitrary unknown quantum state are impossible since the linearity of quantum operations forbids such a replication \cite{no-deleting,Wootters,Dieks}.

Quantum no-cloning theorem was found by Wootters, Zurek, and Dieks in 1982 \cite{Wootters,Dieks}, and has profound implications in quantum computing and related fields \cite{review1,review2}. It can be proved in two different ways for quantum no-cloning theorem. One is by using the linearity of quantum mechanics, which is first proposed in \cite{Wootters,Dieks}. The other one is by using the properties of unitary operation, which is first presented by Yuen \cite{Yuen}.
The second proof of the no-cloning theorem in conventional quantum mechanics indicates that two different and nonorthogonal states cannot be perfectly and deterministically cloned. Furthermore, Duan and Guo proposed a different quantum cloning strategy: perfect cloning of linearly independent quantum states (e.g. two different and nonorthogonal states) can be achieved only with a success probability $p$ that is less than unity. This kind of quantum cloning is called probabilistic quantum cloning \cite{probabilistic}.

The quantum no-deleting theorem is first introduced by Pati and Braunstein \cite{no-deleting}, which states that it is impossible to delete an unknown quantum state $|\psi\rangle$ against a copy, i.e.,
\begin{eqnarray}
U_{del}|\psi\rangle|\psi\rangle|A\rangle=|\psi\rangle|\Sigma\rangle|A_{\psi}\rangle,
\end{eqnarray}
the above unitary  operation $U_{del}$ to delete one of two copies of  $|\psi\rangle$ does not exist, where $|\psi\rangle$ is an arbitrary unknown qubit state, $|A\rangle$ is the initial state of ancilla, $|A_{\psi}\rangle$ is the final state of ancilla which may depend on $|\psi\rangle$, and $|\Sigma\rangle$ is some standard final state of the qubit. The quantum no-deleting theorem indicates that  one cannot delete one of two copies of two different and nonorthogonal quantum states by the same unitary operation.

It is well known for a long time that some non-Hermitian Hamiltonians have real eigenvalues. For example, Bender and his collaborators introduced a class of non-Hermitian Hamiltonians with $\mathcal{PT}$-symmetry, which have infinite, discrete, and entirely real and positive spectrum \cite{Bender1,Bender2,Bender3}: $H=p^2-(ix)^N$,
where $N$ is a real number and not less than 2. Recently, $\mathcal{PT}$-symmetry has been observed in classical optics \cite{PT_optic1,PT_optic2,PT_optic3,PT_optic4,PT_optic5,PT_optic6,PT_optic7,PT_optic8,PT_optic9,PT_optic10,PT_optic11,PT_optic12,PT_optic13,PT_optic14}, and non-Hermitian Hamiltonian with real eigenvalues has attracted much interest \cite{PT1,PT2,PT5,PT6,PT7,exp1,exp2,Ali1,Ali2,Ali3,Ali4,PT3,PT4,comment1,comment2}. In the quantum regime, some experimental results have been reported to simulate non-Hermitian system in an open conventional quantum system \cite{PT7,exp1,exp2}. For example, an experimental investigation has been reported by using an open quantum system to simulate the $\mathcal{PT}$-symmetric system as a part of the full Hermitian system \cite{PT7}.

In this work, the quantum deleting and cloning will be investigated in a pseudo-unitary system. We first present a pseudo-Hermitian Hamiltonian with real eigenvalues in a two-qubit system, which has recently been reported in an experimental simulation \cite{exp2}. Here, we extend its theoretical part. By using the pseudo-unitary operators generated from this pseudo-Hermitian Hamiltonian, one can show that it is possible to delete and clone a class of two different and nonorthogonal states, and it can be generalized to arbitrary two different and nonorthogonal pure qubit states. The cloning has already been simulated in the experiment \cite{exp2}. Furthermore, quantum state discrimination, which is strongly related to quantum no-cloning theorem, will also be discussed. Last but not least, we simulate these pseudo-unitary pseudo-unitary operators in conventional quantum mechanics with post-selection, and obtain the success probability of simulations. Pseudo-unitary operators are implemented with a limited efficiency due to the post-selections. Thus, the success probabilities of deleting and cloning in the simulation by conventional quantum  mechanics are less than unity,  which does not violate quantum no-deleting and no-cloning theorem. Investigating quantum deleting and cloning in a pseudo-Hermitian system will provide insight into the physical content and mathematical structure of pseudo-Hermitian quantum mechanics.


\section{Pseudo-Hermitian Hamiltonian with real spectrum}
We focus on one kind of non-Hermitian Hamiltonian, the pseudo-Hermitian Hamiltonian. Recall that a Hamiltonian $\mathcal{H}$ is pseudo-Hermitian if and only if there exists a metric operator $\eta$ which has its inverse matrix $\eta^{-1}$, such that \cite{Ali1,Ali2,Ali3}
\begin{equation}\label{p_H}
    \mathcal{H}^\dag=\eta \mathcal{H} \eta^{-1}.
\end{equation}
The pseudo-Hermitian can be viewed as a natural generalization from Hermitian, since when $\eta=\idol$ pseudo-Hermitian reduces to Hermitian. The metric operator $\eta$, which is usually Hermitian and invertible, defines a new inner product \cite{Ali1,Ali2,Ali3},
\begin{equation}\label{inner}
    \langle\psi_1|\psi_2\rangle_{\eta}:=\langle\psi_1|\eta|\psi_2\rangle.
\end{equation}

For a pseudo-Hermitian Hamiltonian $\mathcal{H}$ with real eigenvalues, we can see $\mathcal{H}\neq \mathcal{H}^\dag$ if $\eta\neq \idol$. So the pseudo-unitary operator reads \cite{Ali1,Ali2,Ali3}
\begin{eqnarray}
\mathcal{U}(t)=e^{-i\mathcal{H}t/\hbar},
\end{eqnarray}
without loss of generality we assume $\hbar=1$. In a direct consequence, it can be found that if $\eta\neq \idol$ then
\begin{eqnarray}
\mathcal{U}^\dag(t)\mathcal{U}(t)=e^{i\mathcal{H}^\dag t}e^{-i\mathcal{H} t}\neq \idol.
\end{eqnarray}
Instead, $\mathcal{U}(t)$ is pseudo-unitary satisfying \cite{Ali1,Ali2,Ali3}
\begin{eqnarray}
\mathcal{U}^\dag(t)\eta\mathcal{U}(t)=\eta,
\end{eqnarray}
since $\mathcal{H}$ is a pseudo-Hermitian Hamiltonian. Thus, one may find that for some special states, the cloning of two different and nonorthogonal states is possible by pseudo-unitary operators.

We first review a two-qubit pseudo-Hermitian Hamiltonian with every eigenvalue being real, which has been proposed in Ref. \cite{exp2},
\begin{eqnarray}
\mathcal{H}&=&\frac{1}{4}\big(a\sigma_z-i b \sigma_y\big)\otimes\big(2\idol-\sqrt{2}\sigma_x-c^* \sigma_y-d^* \sigma_z\big)\nonumber\\
&&+\frac{1}{4}\idol\otimes\big(2\idol-\sqrt{2}\sigma_x+c \sigma_y-d \sigma_z\big),\label{H}
\end{eqnarray}
where $\idol$, $\sigma_x$, $\sigma_y$ and $\sigma_z$ are identity and Pauli matrices, $a=\cosh{\theta}$, $b=\sinh{\theta}$, $c=\cosh{\theta}+i\sqrt{2}\sinh{\theta}$, $d=\sqrt{2}\cosh{\theta}-i\sinh{\theta}$. The (right) eigenvalues, $E=1/2,-1/2,2,0$, are real, corresponding to the (right) eigenstates,
\begin{eqnarray}
&&|E_1\rangle=\frac{1}{\sqrt{2}}\left( \begin{array}{c}
a-i b-1\\
-i(1+a+b) \\
i(1-a)+b\\
a-i b+1
\end{array} \right),\\
&&|E_2\rangle=\frac{1}{\sqrt{2}}\left( \begin{array}{c}
a+i b-1\\
i(1+a+b) \\
i(a-1)+b\\
a+i b+1
\end{array} \right),\\
&&|E_3\rangle=\frac{1}{4 - 2 \sqrt{2}}\left( \begin{array}{c}
(1-\sqrt{2})(a+1)+b \\
a+(1-\sqrt{2})b-1 \\
a+(1-\sqrt{2})b+1\\
(1-\sqrt{2})(a-1)+b
\end{array} \right),\\
&&|E_4\rangle=\frac{1}{4 + 2 \sqrt{2}}\left( \begin{array}{c}
(1+\sqrt{2})(a+1)+b \\
a+(1+\sqrt{2})b-1 \\
a+(1+\sqrt{2})b+1\\
(1+\sqrt{2})(a-1)+b
\end{array} \right).
\end{eqnarray}
These eigenstates are not orthogonal to each other in conventional quantum theory. The pseudo-Hermitian Hamiltonian Eq.~(\ref{H}) actually belongs to pseudo-Hermitian Hamiltonians \cite{Ali1,Ali2,Ali3}, i.e., satisfying Eq. (\ref{p_H}) with
\begin{equation}\label{eta}
  \eta=\eta_0^{\otimes 2}=(a\idol-b\sigma_x)^{\otimes 2},
\end{equation}
and  the (right) eigenstates $\{|E_i\rangle\}$ satisfy
\begin{equation}\label{}
  \langle E_i|E_j\rangle_\eta=\langle E_i|\eta|E_j\rangle=\delta_{ij},
\end{equation}
i.e.,  these (right) eigenstates $\{|E_i\rangle\}$ are orthogonal to each other and normalized under the new inner product defined by $\eta$.
Furthermore, similar to the conventional quantum mechanics, the pseudo-unitary operator for such a system is given by
\begin{eqnarray}
\mathcal{U}(t)\equiv e^{-i\mathcal{H}t},\label{U}
\end{eqnarray}
where we assume $\hbar=1$.

\section{Quantum deleting with pseudo-unitary operator}
The quantum no-deleting theorem is first introduced by Pati and Braunstein \cite{no-deleting}, which states that  one cannot delete one of two copies of two different and nonorthogonal quantum states by the same unitary operation. The main idea is to choose a special pseudo-Hermitian Hamiltonian such that two different and nonorthogonal states in conventional quantum mechanics can be orthogonal with respect to the new definition of inner product determined by that pseudo-Hermitian Hamiltonian. To simulate this pseudo-unitary operator in conventional quantum mechanics, one needs post-selections which lead to the success probability of deleting less than unity.

Suppose there exists a deleting machine which can perfectly delete two different and nonorthogonal quantum states with two copies $|\alpha_1\rangle_A\otimes|\alpha_1\rangle_B$ and $|\alpha_2\rangle_A\otimes|\alpha_2\rangle_B$ in system $AB$. In order to make deleting, one needs to obtain the blank state $|+\rangle_B=(|0\rangle+|1\rangle)/\sqrt{2}$ in system $B$ after deleting. It is worth noticing that the blank state $|+\rangle_B$ is independent of the choice of $|\alpha_1\rangle$ and $|\alpha_2\rangle$ in system $A$, since we have no prior knowledge of $|\alpha_1\rangle$ or $|\alpha_2\rangle$ in system $A$. Thus, the final states after deleting are then described by the tensor product, i.e., $|\alpha_i\rangle_A \otimes |+\rangle_B$ with $i=1,2$.

In conventional quantum systems, one cannot delete one of two copies of two different and nonorthogonal quantum states by the same unitary operation $U_{del}$. However, in pseudo-Hermitian systems, it is possible. Suppose there exists a perfect deleting machine which can perfectly delete two different and nonorthogonal quantum states $|\alpha_1\rangle$ and $|\alpha_2\rangle$. We choose the pseudo-Hermitian Hamiltonian as
\begin{eqnarray}
\mathcal{H}'=-\mathcal{H},\label{H'}
\end{eqnarray}
where $\mathcal{H}$ is present in Eq. (\ref{H}). The (right) eigenvalues of $\mathcal{H}'$ are real, i.e., $E'=-1/2,1/2,-2,0$.
In order to perfectly delete $|\alpha_1\rangle$ and $|\alpha_2\rangle$, we set the time of evolution to $\tau=\pi/2$ in Eq. (\ref{U}). Therefore, the pseudo-unitary  operator $\mathcal{U}_{del}(\tau)$ is as follows,
\begin{eqnarray}
&&\mathcal{U}_{del}(\tau)=e^{-i\mathcal{H}'\pi/2}\nonumber\\
&=&\frac{1}{2\sqrt{2}}\big(a\sigma_z-i b \sigma_y\big)\otimes\Big(-\idol+\sigma_x+(a+b)(\sigma_z-i\sigma_y)\Big)\nonumber\\
&&+\frac{1}{2\sqrt{2}}\idol\otimes\Big(\idol+\sigma_x+(a-b)(\sigma_z+i\sigma_y)\Big).\label{delete}
\end{eqnarray}
Consider the process of deleting machine as the pseudo-unitary operator $\mathcal{U}_{del}(\tau)$, then we have
\begin{eqnarray}
\mathcal{U}_{del}(\tau)|\alpha_1\rangle_A|\alpha_1\rangle_B&\propto&|\alpha_1\rangle_A|+\rangle_B,\\
\mathcal{U}_{del}(\tau)|\alpha_2\rangle_A|\alpha_2\rangle_B&\propto&|\alpha_2\rangle_A|+\rangle_B,
\end{eqnarray}
where the two different and nonorthogonal quantum states $|\alpha_1\rangle$ and $|\alpha_2\rangle$ are as  follows \cite{exp2},
\begin{eqnarray}
&&|\alpha_1\rangle=\frac{1}{\sqrt{\cosh{\theta}}}\left( \begin{array}{c}
\cosh{\frac{\theta}{2}}\\
\sinh{\frac{\theta}{2}}
\end{array} \right),\label{alpha1}\\
&&|\alpha_2\rangle=\frac{1}{\sqrt{\cosh{\theta}}}\left( \begin{array}{c}
\sinh{\frac{\theta}{2}}\\
\cosh{\frac{\theta}{2}}
\end{array} \right).\label{alpha2}
\end{eqnarray}
One can easily check \cite{exp2}
\begin{eqnarray}
\langle\alpha_1|\alpha_2\rangle&=&\tanh{\theta},\\
\langle\alpha_1|\alpha_2\rangle_{\eta_0}&=&\langle\alpha_1|\eta_0|\alpha_2\rangle=0,
\end{eqnarray}
with $\eta_0$ being given in Eq. (\ref{eta}), i.e., $|\alpha_1\rangle$ and $|\alpha_2\rangle$ are not orthogonal in conventional quantum mechanics, but they are orthogonal under the new defined inner product. Thus, the pseudo-unitary operator $\mathcal{U}_{del}(\tau)$ coming from the pseudo-Hermitian Hamiltonian Eq. (\ref{H'}) can be used to  perfectly delete two different and nonorthogonal quantum states $|\alpha_1\rangle$ and $|\alpha_2\rangle$, since $|\alpha_1\rangle$ and $|\alpha_2\rangle$ are orthogonal under the new inner product defined by $\eta_0$.

Our approach works for arbitrary two different and nonorthogonal pure qubit states. Consider two arbitrary different and nonorthogonal pure qubit states $|\psi_1\rangle$ and $|\psi_2\rangle$, there always exists an unitary matrix $V$ and $\theta$ such that $V|\psi_1\rangle=|\alpha_1\rangle$ and $V|\psi_2\rangle=|\alpha_1\rangle$ \cite{probabilistic}. Thus, we can construct a new pseudo-Hermitian Hamiltonian $\mathcal{H}'_V$ and its time-evolution operator $\mathcal{U}_V(\tau)$ from Eqs. (\ref{H'}) and (\ref{delete}),
\begin{eqnarray}
\mathcal{H}'_V&=&V^\dag\otimes V^\dag \mathcal{H}' V\otimes V,\\
\mathcal{U}_V(\tau)&=&V^\dag\otimes V^\dag \mathcal{U}_{del}(\tau) V\otimes V.
\end{eqnarray}
Therefore, one has
\begin{eqnarray}
\mathcal{U}_V(\tau)|\psi_1\rangle_A|\psi_1\rangle_B&\propto&|\psi_1\rangle_A|+'\rangle_B,\\
\mathcal{U}_V(\tau)|\psi_2\rangle_A|\psi_2\rangle_B&\propto&|\psi_2\rangle_A|+'\rangle_B,
\end{eqnarray}
where $|+'\rangle_B=V^\dag|+\rangle_B$, and $|+'\rangle$ is still independent of the choice of $|\psi_1\rangle$ and $|\psi_2\rangle$ in system $A$.

\section{Quantum cloning with pseudo-unitary operator}
Here, for the sake of integrity, we review quantum cloning in a pseudo-unitary system, which has already been experimentally simulated in Ref. \cite{exp2}.

In order to perfectly clone $|\alpha_1\rangle$ and $|\alpha_2\rangle$, we set the time of evolution to $\tau=\pi/2$ in Eq. (\ref{U}). Therefore, the pseudo-unitary operator $\mathcal{U}_{clone}(\tau)$ is as follows \cite{exp2},
\begin{eqnarray}
&&\mathcal{U}_{clone}(\tau)=e^{-i\mathcal{H}\pi/2}\nonumber\\
&=&\frac{1}{2\sqrt{2}}\big(a\sigma_z-i b \sigma_y\big)\otimes\Big(-\idol+\sigma_x+(a-b)(\sigma_z+i\sigma_y)\Big)\nonumber\\
&&+\frac{1}{2\sqrt{2}}\idol\otimes\Big(\idol+\sigma_x+(a+b)(\sigma_z-i\sigma_y)\Big).\label{Utau}
\end{eqnarray}
Consider the process of cloning machine as the pseudo-unitary operator $\mathcal{U}_{clone}(\tau)$, then we have \cite{exp2}
\begin{eqnarray}
\mathcal{U}_{clone}(\tau)|\alpha_1\rangle_A|+\rangle_B&\propto&|\alpha_1\rangle_A|\alpha_1\rangle_B,\\
\mathcal{U}_{clone}(\tau)|\alpha_2\rangle_A|+\rangle_B&\propto&|\alpha_2\rangle_A|\alpha_2\rangle_B,
\end{eqnarray}
where the two different and nonorthogonal quantum states $|\alpha_1\rangle$ and $|\alpha_2\rangle$ are as Eqs. (\ref{alpha1}) and (\ref{alpha2}). One can easily check $\langle\alpha_1|\alpha_2\rangle=\tanh{\theta}$. Thus, the pseudo-unitary  operator $\mathcal{U}_{clone}(\tau)$ coming from the pseudo-Hermitian Hamiltonian Eq. (\ref{H}) can be used to  perfectly clone two different and nonorthogonal quantum states $|\alpha_1\rangle$ and $|\alpha_2\rangle$, since $|\alpha_1\rangle$ and $|\alpha_2\rangle$ are orthogonal under the new inner product defined by $\eta_0$.

\section{State discrimination}
In conventional quantum mechanics, the state discrimination is a well-known problem, in which we would like to distinguish optimally between a given set of quantum states with one measurement. For instance, suppose that the state is known to be one of two possible pure states $|\psi_1\rangle$ and $|\psi_2\rangle$ with associated probabilities $p_1$ and $p_2$ ($p_1+p_2=1$), respectively. If $|\psi_1\rangle$ and $|\psi_2\rangle$ are orthogonal, one can perfectly determine the state by choosing the project measurements $\hat{\pi}_1=|\psi_1\rangle\langle\psi_1|$ and $\hat{\pi}_2=|\psi_2\rangle\langle\psi_2|$ ($\hat{\pi}_1+\hat{\pi}_2=\idol$). However, if $|\psi_1\rangle$ and $|\psi_2\rangle$ are not orthogonal, there is a strategy in which successful state discrimination can be achieved with a single measurement but only with a success probability
$p_{succ}$ that is less than unity. The minimum probability of making an error is given by the so-called Helstrom bound \cite{Helstrom,Barnett},
\begin{eqnarray}
p_{err}=\frac{1}{2}(1-\sqrt{1-4p_1p_2|\langle\psi_1|\psi_2\rangle|^2}),
\end{eqnarray}
which minimizes the error probability associated with a single measurement.

However, in pseudo-Hermitian Hamiltonian systems, one may perfectly clone the two nonorthogonal states Eqs. (\ref{alpha1}) and (\ref{alpha2}). Without perfectly cloning,  the minimum error probability for $|\alpha_1\rangle$ in Eq. (\ref{alpha1}) and $|\alpha_2\rangle$ in Eq. (\ref{alpha2}) would be
\begin{equation}\label{}
   p_{err}'=\frac{1}{2}(1-\sqrt{1-4p_1p_2\tanh^2{\theta}}).
\end{equation}
If one used $\mathcal{U}_{clone}(\tau)$ in Eq. (\ref{Utau}) for $N-1$ times, the two possible pure states become to $|\alpha_1\rangle^{\otimes N}$ and $|\alpha_2\rangle^{\otimes N}$, and distinguishing between $|\alpha_1\rangle$ and $|\alpha_2\rangle$ is equivalent to distinguishing between $|\alpha_1\rangle^{\otimes N}$ and $|\alpha_2\rangle^{\otimes N}$. Therefore, the Helstrom bound for $|\alpha_1\rangle^{\otimes N}$ and $|\alpha_2\rangle^{\otimes N}$ would be
\begin{equation}\label{}
   p_{err}''=\frac{1}{2}(1-\sqrt{1-4p_1p_2\tanh^{2N}{\theta}}),
\end{equation}
where $p_{err}''< p_{err}'$ since $|\tanh{\theta}|<1$. Furthermore, one has $p_{err}''\rightarrow0$ when $N\rightarrow +\infty$, which means one can almost perfectly distinguish between $|\alpha_1\rangle$ and $|\alpha_2\rangle$ with infinity copies.

\section{Simulation in conventional quantum mechanics with post-selection}
Actually, the pseudo-unitary  operators $\mathcal{U}(\tau)$ in Eqs. (\ref{delete}) and (\ref{Utau}) can be effectively simulated by a conventional quantum gate with post-selection. This is very similar to the cases in Refs. \cite{PT7,PT3,PT4}. Since the evolution of $\mathcal{U}(\tau)$ is pseudo-unitary, an ancilla and a post-selecting projection operation are necessary. Suppose that the initial state of the system is $|\psi_I\rangle$. To realize $\mathcal{U}(\tau)$ in conventional quantum mechanics, one needs an ancilla qubit $|0\rangle$. Therefore, the initial state of the total system, including the ancilla qubit, is $|\psi_I\rangle\otimes|0\rangle$.

Suppose that the singular value decomposition of $\mathcal{U}(\tau)$ is $\mathcal{U}(\tau)=U\Sigma V^\dag$, where $U$ and $V$ are unitary matrices, and $\Sigma=\mathrm{diag}\{\lambda_1,\lambda_2,\lambda_3,\lambda_4\}$ is a diagonal matrix with $\{\lambda_i\}$ being singular values of $\mathcal{U}(\tau)$ \cite{horn}. There are two situations as follows.

a) If the maximal singular value $\lambda_{max}:=\max\{\lambda_1,\lambda_2,\lambda_3,\lambda_4\}$ is greater than 1, we define a normalized $\mathcal{U}(\tau)$ as
\begin{equation}\label{}
  \widetilde{\mathcal{U}}(\tau):= \frac{\mathcal{U}(\tau)}{\lambda_{max}},
\end{equation}
and construct a new matrix
\begin{equation}\label{}
  \mathcal{V}:=U\Sigma' V^\dag
\end{equation}
with
\begin{equation}\label{}
  \Sigma'=\mathrm{diag}\{\lambda_1',\lambda_2',\lambda_3',\lambda_4'\}
\end{equation}
and
\begin{equation}\label{}
  \lambda_i'=\sqrt{1-\frac{\lambda_i^2}{\lambda_{max}^2}}.
\end{equation}

b) If $\lambda_{max}\leq1$, then we define
\begin{eqnarray}
\widetilde{\mathcal{U}}(\tau)&:=&\mathcal{U}(\tau)  \\
\mathcal{V}&:=&U\tilde{\Sigma}' V^\dag        \\
\end{eqnarray}
with
\begin{equation}\label{}
\tilde{\Sigma}'=\mathrm{diag}\{\tilde{\lambda}_1',\tilde{\lambda}_2',\tilde{\lambda}_3',\tilde{\lambda}_4'\}
\end{equation}
and
\begin{equation}\label{}
  \tilde{\lambda}_i'=\sqrt{1-\lambda_i^2}.
\end{equation}

It is worth noticing that for the above two cases we can always prove
\begin{eqnarray}\label{}
  \widetilde{\mathcal{U}}(\tau)\widetilde{\mathcal{U}}(\tau)^\dag+\mathcal{V}\mathcal{V}^\dag&=&\idol, \label{cond1} \\
  \widetilde{\mathcal{U}}(\tau)^\dag\widetilde{\mathcal{U}}(\tau)+\mathcal{V}^\dag\mathcal{V}&=&\idol, \label{cond2} \\
  \widetilde{\mathcal{U}}(\tau)\mathcal{V}^\dag-\mathcal{V}\widetilde{\mathcal{U}}^\dag&=&0, \label{cond3}   \\
\mathcal{V}^\dag\widetilde{\mathcal{U}}(\tau)-\widetilde{\mathcal{U}}^\dag\mathcal{V}&=&0.\label{cond4}
\end{eqnarray}
Thus, one can construct an unitary evolution of the total system, which can be expressed as
\begin{eqnarray}
U_{tot}&=&\widetilde{\mathcal{U}}(\tau)\otimes|0\rangle\langle0|+\mathcal{V}\otimes|1\rangle\langle0|\nonumber\\
&&-\mathcal{V}\otimes|0\rangle\langle1|+\widetilde{\mathcal{U}}(\tau)\otimes|1\rangle\langle1|,
\end{eqnarray}
where $\widetilde{\mathcal{U}}(\tau)$ and $\mathcal{V}$ are defined as above. Based on Eqs. (\ref{cond1}-\ref{cond4}), we can easily check that the unitary evolution of the total system $U_{tot}$ is unitary, i.e.,
\begin{equation}\label{}
  U_{tot}U_{tot}^\dag=U_{tot}^\dag U_{tot}=\idol,
\end{equation}
which means $U_{tot}$ can be realized in conventional quantum mechanics.
Then the initial state $|\psi_I\rangle\otimes|0\rangle$  after the unitary evolution $U_{tot}$ is   $U_{tot}|\psi_I\rangle\otimes|0\rangle$.
We choose the projection operator on the ancilla qubit as $P=\idol\otimes|0\rangle\langle0|$. Thus, the evolution after the projection operator (or post-selection) can be calculated as
\begin{eqnarray}
PU_{tot}|\psi_I\rangle\otimes|0\rangle=\Big(\widetilde{\mathcal{U}}(\tau)|\psi_I\rangle\Big)\otimes|0\rangle.
\end{eqnarray}
So for an arbitrary  initial state of the system $|\psi_I\rangle$, the evolution after the projection operator (or post-selection) is $\mathcal{U}_{ps}=\widetilde{\mathcal{U}}(\tau)\propto \mathcal{U}(\tau)$.

In the following, we will calculate the success  probability of the above simulation. It is worth noticing that the initial state of the system $|\psi_I\rangle$ is normalized in  conventional quantum mechanics, i.e., $\langle \psi_I|\psi_I\rangle=1$, and $U_{tot}|\psi_I\rangle\otimes|0\rangle$ is still normalized in  conventional quantum mechanics, since $U_{tot}$ is unitary. Thus,
\begin{eqnarray}
&&U_{tot}|\psi_I\rangle\otimes|0\rangle  \nonumber\\
&=&\widetilde{\mathcal{U}}(\tau)|\psi_I\rangle\otimes|0\rangle+\mathcal{V}|\psi_I\rangle\otimes|1\rangle  \nonumber\\
&=&\sqrt{N_1}\frac{\widetilde{\mathcal{U}}(\tau)|\psi_I\rangle}{\sqrt{N_1}}\otimes|0\rangle+\sqrt{N_2}\frac{\mathcal{V}|\psi_I\rangle}{\sqrt{N_2}}\otimes|1\rangle, \label{Utot}
\end{eqnarray}
where $N_1:=\langle \psi_I|\widetilde{\mathcal{U}}(\tau)^\dag \widetilde{\mathcal{U}}(\tau)|\psi_I\rangle$ and $N_2:=\langle\psi_I|\mathcal{V}^\dag \mathcal{V}|\psi\rangle$ are the normalization factors. Moreover, from Eq. (\ref{cond2}), we can see that
\begin{equation}\label{}
  N_1+N_2=\langle \psi_I|\big(\widetilde{\mathcal{U}}(\tau)^\dag \widetilde{\mathcal{U}}(\tau)+\mathcal{V}^\dag \mathcal{V}\big)|\psi_I\rangle=1.
\end{equation}
Based on Eq. (\ref{Utot}), one can get the success  probability $p$  after  the projection operator $P=\idol\otimes|0\rangle\langle0|$ on the ancilla qubit as
\begin{eqnarray}
p&=&\langle \psi_I| \otimes \langle 0| U_{tot}^\dag  P U_{tot}|\psi_I\rangle  \otimes  |0\rangle  \nonumber\\
&=& \langle \psi_I|\widetilde{\mathcal{U}}(\tau)^\dag \widetilde{\mathcal{U}}(\tau)|\psi_I\rangle \nonumber\\
&=& N_1.
\end{eqnarray}
On the other hand, the probability of failure is $1-p=N_2$, and in this cas one can only get $\mathcal{V}|\psi_I\rangle$ rather than $\widetilde{\mathcal{U}}(\tau)|\psi_I\rangle$.

Therefore, a simulation experiment can, in principle, be realized by conventional quantum mechanics with post-selection. Since the post-selection is necessary in the conventional quantum mechanics simulation, the  success probability is less than unity. Thus, this simulation can be regarded as probabilistic quantum deleting and cloning.

\section{Discussions and conclusions}
Actually, Ref. \cite{PT5} has already discussed the state discrimination problem in $\mathcal{PT}$-symmetry systems.
The differences and novelties between this work and Ref. \cite{PT5} are as follows:
a) In Ref. \cite{PT5}, the used Hamiltonian is $\mathcal{PT}$-symmetric with real eigenvalues; but in this work, the Hamiltonian we used is pseudo-Hermitian without $\mathcal{PT}$ symmetry. In general, finite-dimensional $\mathcal{PT}$-symmetric Hamiltonian with real eigenvalues must be pseudo-Hermitian, but conversely it may not be true.
b) In Ref. \cite{PT5}, they proposed two different but related solutions to the non-Hermitian simulated unambiguous state-discrimination problem. In this work, we propose another different solution by using the quantum cloning.
c) In this work, we have presented a general method to simulate pseudo-unitary  operators  in conventional quantum mechanics with post-selection, which has not being involved in Ref. \cite{PT5}.

The experimental simulation of quantum cloning  has been demonstrated in Ref. [31]. One may want to know the difference for the experimental realizations on quantum deletion and quantum cloning. The differences are as follows: a) The initial states are different. For the experimental realization on quantum deleting, one needs to prepare $|\alpha_i\rangle_A|\alpha_i\rangle_B$ ($i=1,2$) as the initial state, but for the experimental realization on quantum cloning, the initial state should be $|\alpha_i\rangle_A|+\rangle_B$ ($i=1,2$). b) The pseudo-unitary operators are different. The pseudo-unitary operator for quantum deleting is $\mathcal{U}_{del}(\tau)=e^{-i\mathcal{H}'\pi/2}$, but for quantum cloning it is $\mathcal{U}_{clone}(\tau)=e^{-i\mathcal{H}\pi/2}$. c) The simulations for pseudo-unitary operators $\mathcal{U}_{del}(\tau)$ and $\mathcal{U}_{clone}(\tau)$ in conventional quantum mechanics with post-selection are different.

Investigation of quantum deleting and cloning in a pseudo-unitary system is an interesting topic. It allows us to imagine another universe which has almost the same quantum mechanics but with a different inner product. Furthermore, our work presents a general method to simulate pseudo-unitary operators in that universe, and  provides a platform to study pseudo-unitary operators in conventional quantum mechanics.

\section*{ACKNOWLEDGMENTS}
We are grateful to Franco Nori, Adam Miranowicz, Dorje C. Brody, Carl M. Bender, Otfried G\"uhne, Ray-Kuang Lee, Qing Chen, Sixia Yu and Yong-Jian Han for valuable discussions.
This work is funded by the National Natural Science Foundation of China (China) (Grants No. 11734015, 11474049, 11674056), the K.C. Wong Magna Fund in Ningbo University, the financial support from Research Grants Council of Hong Kong (RGC, Hong Kong) (Grant No. 538213). M.G. is supported by the National Youth Thousand Talents Program (Grant No. KJ2030000001), the USTC start-up funding (Grant No. KY2030000053).

\end{document}